\documentclass[sigconf]{acmart}

\AtBeginDocument{%
  \providecommand\BibTeX{{%
    \normalfont B\kern-0.5em{\scshape i\kern-0.25em b}\kern-0.8em\TeX}}}

\copyrightyear{2023}
\acmYear{2023}
\setcopyright{acmlicensed}
\acmConference[ESEM '23]{International Symposium on Empirical Software Engineering and Measurement - Doctoral Symposium Track}{October 22--27, 2023}{New Orleans, Louisiana, United States}

\acmPrice{15.00}
\acmDOI{}
\acmISBN{}

\usepackage{array}
\usepackage{multirow}

\usepackage{xspace}

\usepackage[normalem]{ulem} 
\usepackage{xcolor}

\newcommand{\del}[1]{} 

\begin{document}

\title{
Learning from Interaction: User Interface Adaptation using Reinforcement Learning
}

\author{Daniel Gaspar-Figueiredo}
\affiliation{%
  \institution{Universitat Politècnica de València \& Instituto Tecnológico de Informática }
  \city{Valencia}
  \country{Spain}
  \postcode{--}}
\email{dagasfi@epsa.upv.es}

\renewcommand{\shortauthors}{Daniel Gaspar-Figueiredo}

\begin{abstract}

The continuous adaptation of software systems to meet the evolving needs of users is very important for enhancing user experience (UX). User interface (UI) adaptation, which involves adjusting the layout, navigation, and content presentation based on user preferences and contextual conditions, plays an important role in achieving this goal. 
However, suggesting the right adaptation at the right time and in the right place remains a challenge in order to make it valuable for the end-user.
To tackle this challenge, machine learning approaches could be used. In particular, we are using Reinforcement Learning (RL) due to its ability to learn from interaction with the users. In this approach, the feedback is very important and the use of physiological data could be benefitial to obtain objective insights into how users are reacting to the different adaptations. Thus, in this PhD thesis, we propose an RL-based UI adaptation framework that uses physiological data.
The framework aims to learn from user interactions and make informed adaptations to improve UX. To this end, our research aims to answer the following questions: \textit{Does the use of an RL-based approach improve UX?} \textit{How effective is RL in guiding UI adaptation?} and \textit{Can physiological data support UI adaptation for enhancing UX?} 
The evaluation plan involves conducting user studies to evaluate answer these questions. The empirical evaluation will provide a strong empirical foundation for building, evaluating, and improving the proposed adaptation framework.
The expected contributions of this research include the development of a novel framework for intelligent Adaptive UIs, insights into the effectiveness of RL algorithms in guiding UI adaptation, the integration of physiological data as objective measures of UX, and empirical validation of the proposed framework's impact on UX.

\end{abstract}



\maketitle

\raggedbottom

\section{Introduction}
\label{sec:intro}
Software development approaches that effectively respond to the needs of end-users have gained significant attention. Among these approaches, adaptive systems have emerged as a promising solution for designing software systems that can automatically adjust their behavior and presentation based on changing contextual conditions. This adaptability aims to enhance the User Experience (UX) and improve overall user satisfaction.

One specific area where adaptive systems play an important role is user interface adaptation~\cite{Calvary:2003}. The user interface (UI) comprises various elements, including classical widgets, custom widgets, text, and graphics, which enable users to interact with the system. Adaptive User Interfaces (AUIs)~\cite{Findlater:2009} have been introduced to address the evolving needs of users over time and deliver a personalized experience by adjusting the layout, navigation, and content presentation in real-time based on each user's preferences and abilities. 

To this end, machine learning (ML) techniques can be used in order to automate and optimize the adaptation process in AUIs. These algorithms can analyze patterns, preferences, and contextual information to make intelligent decisions regarding interface adaptation~\cite{abrahaoModel:2021}.
One Machine Learning technique that shows a great potential in guiding the adaptation process is Reinforcement Learning (RL). RL has gained attention for its ability to learn optimal decision-making policies through interactions with an environment. In the context of AUIs, RL allows the adaptive interfaces to continuously learn and improve their adaptation strategies based on trial and error interactions with an environment and receiving feedback in the form of rewards~\cite{sutton:2018, Chen:2022} by actively exploring different adaptation strategies and exploiting the ones that yield positive outcomes. The ability of RL to consider long-term goals and optimize cumulative rewards over time provides a robust framework for guiding the adaptation process and improving user experience. This technique enables AUIs to adapt to dynamic user contexts, preferences, and evolving UX requirements.

Understanding the user emotional state can be powerful to detect the evolving UX requirements.
This is typically assessed using retrospective methods (e.g., interviews and questionnaires), thus increasing the bias of self-report~\cite{Vermeeren:2010} and the Hawthorne effect~\cite{McCarney:2007}. Since these methods do not capture the UX during the real interaction, other approaches have been receiving an increasing interest, such as  physiological data~\cite{Graziotin:2022}. 
It can be captured from sensors and human physiological responses, such as electroencephalography, eye tracking, and more. 
It can be used to effectively measure various aspects of the UX in real-time, such as attractiveness, cognitive effort, engagement, among others.
Although these measures are complex to interpret, they are objective \cite{bergstrom2014physiological}.

In this context, this research aims to address the following research question and sub-questions:

\begin{itemize}

    
    \item[RQ] Does the use of an RL-based approach for UI adaptation improve UX?

    \begin{itemize}

        \item[RQ$_{1}$] How effective is RL in guiding UI adaptation for enhancing the user experience?
    
        \item[RQ$_{2}$] Can physiological data be used to support UI adaptation for enhancing the user experience?
    
    \end{itemize}

    
\end{itemize}

By addressing these research questions, this work aims to advance the understanding and application of adaptive UIs to ensure that they adapt to the users' preferences and needs. 
The methodology that is being followed involves a combination of literature review, the design and implementation of the proposed framework, experimentation with academic and industry environments, and releasing the solution. 
The expected contributions of this research include insights into the effectiveness of RL algorithms in guiding UI adaptation, strategies for obtaining non-intrusive user feedback, and understanding the impact of integrating physiological data in the adaptive UI process. 
These contributions aim to enhance the development of software systems that prioritize UX and deliver more effective and user-centric solutions.

\section{Related Work}
\label{sec:relWork}

One relevant theoretical framework in the field of UI adaptation is the 
PDA-LDA cycle, which is structured according to the theory of control perspective~\cite{bouzit:2017}. This cycle consists of three stages: perception (P) of the context before adaptation, decision (D) to adapt, and action (A) taken to adapt. Both the end-user and the system engage in this cycle symmetrically, with each stage being covered to some extent.

In the field of Human-Computer Interaction (HCI), extensive research has been conducted on UI adaptation methods. However, the practical implementation of these methods has not fully supported the complete PDA-LDA cycle. Existing approaches in the HCI community have primarily focused on specific aspects, such as interface design and user preferences, often neglecting the holistic integration of the perception, decision, and action stages.
On the other hand, the Software Engineering (SE) community has made advancements in principles and technologies to support the PDA-LDA cycle on the system side, such as the MAPE-K adaptation loop and models at runtime. However, the perception, decision, and action stages on the end-user side, which are typically addressed in HCI, are often given a secondary role~\cite{abrahaoModel:2021}.


Therefore, there is a need to bridge the gap between HCI and SE by integrating the perception, decision, and action stages of the UI adaptation process on both the end-user and system sides.

Moreover, as motivated in Section \ref{sec:intro}, there is a growing interest in using machine learning techniques, in particular reinforcement learning, to enhance the UI adaptation process. 
For example, Todi et al. ~\cite{Todi:2021MCTS} used Model-based RL to reduce the selection time in different menu designs. They used predictive HCI models to predict rewards for each state during simulations. Since online simulations can be computationally expensive, a pretrained value network was used to directly obtain value estimates for unexplored states. Training data for this neural network was generated using the predictive HCI models. The main drawback is that the authors focused only on one specific UI element (i.e., graphical menus). When interacting with software, there are more than one UI element that play an important role in UX, thus other experiments that consider a more complete UI representation are needed.

Additionally, 
one challenge of RL in this context is determining the reward of each adaptation. 
To address this issue, some approaches have been defined, such as inverse reinforcement learning, generation of rewards based on user interface quantification models \cite{Khaet:2017, Rice:2014, ElBatran:2014} and predictive HCI models. 
%
%
Some authors have already applied these techniques in UI adaptation. For example, in~\cite{langerak:2022marlui} and~\cite{viadmanoy:2023marlmui}, the authors formulated UI adaptation as a multi-agent RL problem, with both a user agent who learns to interact with an UI so as to complete a task and an interface agent who learns UI adaptations to maximise the user agent’s performance. The joint exploration of both agents makes the adaptative UIs goal-agnostic. 
However, these approaches focused on task completion time and number of steps only and did not considered improving the UX. Additionally, we believe that the feedback from real users to guide the adaptation could be considered. When simulating an user as an RL-agent some important information of the real-world might be not represented. To this end, the RL-agent could interact with the real user to obtain feedback, for example, through the use of physiological data, to obtain user emotions, among others which will enrich the representation of the user and its state.



Existing research has provided valuable insights into UI adaptation methods and the utilization of RL techniques. However, challenges remain in integrating the complete PDA-LDA cycle, considering a more comprehensive representation of UI elements, defining appropriate rewards, and incorporating real user feedback. Addressing these challenges will enable us to enhance the user experience through RL-based adaptive UIs.

\section{Research Plan}

\begin{figure*}
    \centering
    \includegraphics[width=0.55\textwidth]{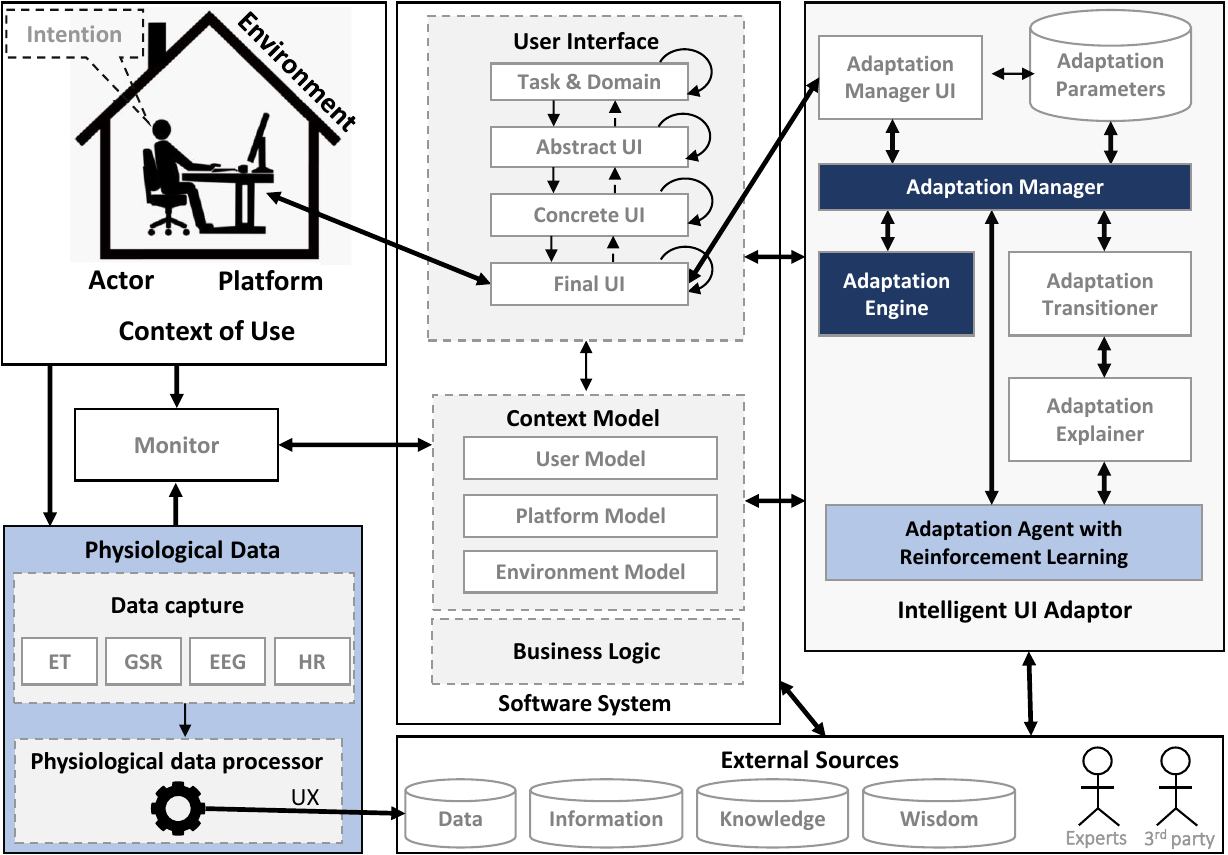}
    \caption{User Interface Adaptation framework using Reinforcement Learning and Physiological data. Dark-blue components are mandatory and light-blue are the main contributions with respect to the original conceptual framework~\cite{abrahaoModel:2021}.} \label{fig:framework}
\end{figure*}

The need for software that continuously adapt to the characteristics of the users, the environment and the platform is forcing developers to consider change as part of the development process. We believe that the current notion of software adaptation will be replaced by one where \textbf{AI and humans collaborate to continuously adapt and evolve software systems to improve the user experience}. This means that humans and the AI-empowered system will collaborate towards a common adaptation goal over time, \textit{evolve together}, improve, and \textit{learn from each other}. 

The initial hypothesis of this PhD thesis is that an approach that enable software adaptations using ML techniques will allow the software industry to improve its productivity, increase the quality of the solutions developed, and at the same time increase the satisfaction of its users. Based on this hypothesis, the general objective of this PhD thesis is to \textit{define and implement a technological framework to support intelligent adaptation of software systems based on a feedback loop fueled by the collaboration between AI and humans}. 

In order to achieve the general objective of this PhD thesis, the following specific objectives are defined:

\begin{itemize}
    \item \textbf{Formalization of the problem of UI adaptation} as a stochastic sequential decision problem. This includes modelling the problem as an Markov Decision Process (MDP).
    \item \textbf{Provide an infrastructure for user modeling} that allows to represent the behavioral and physiological characteristics of users, their preferences, emotions and interaction history. Information related to the platform and the environment will also be represented. These models will provide a knowledge base that will be used to drive the system adaptations. 
     \item \textbf{Provide a repository of adaptation patterns} that will allow to modify the UI with the purpose of improving the UX.
     \item \textbf{Provide a decision-making process based on ML techniques} that will allow exploring the different adaptation alternatives depending on the adaptation goal. This also includes the incorporation of the use of physiological data into the RL adaptation decision-making process and the definition of feedback methods to evaluate the adaptation process. For example through physiological data such as facial expression recognition to obtain the users' emotions.
    \item \textbf{Definition of the environment for RL-based UI adaptation} that will allow to experiment with the proposed framework for intelligent adaptation of software systems.
    \item \textbf{Evaluate the proposed framework and supporting environment} in an academic and industrial context
\end{itemize}

This section outlines our framework that uses RL and physiological data to guide the UI adaptation (Section~\ref{sec:propFramework}) and the evaluation plan (Section~\ref{sec:evaluationPlan}).



\subsection{Proposed Framework}
\label{sec:propFramework}

We propose a framework based on Adaptive Systems principles and predictive Human-Computer Interaction models. The framework incorporates RL, enriched with the use of physiological user data, to address the user interface adaptation problem. This approach leverages the ability of RL to learn and infer knowledge without the need for initial training data. The framework aims to improve the user experience by dynamically adapting the user interface based on contextual factors and user preferences. In this section, we present the framework~\cite{abrahaoModel:2021, gaspar-figueiredo:2022} that we are developing using OpenAI Gym as the RL agent training environment for UI adaptation.

The Figure~\ref{fig:framework} shows the five main parts of the framework with their components:
\begin{itemize}

    \item \textbf{Context of Use}: is the actual conditions under which a given software product is used. It includes factors such as the physical environment, social setting, user characteristics, device characteristics, and any other relevant conditions that may impact the user's interaction with the UI.

    \item \textbf{Software System}: comprises the specification of the system, the user interface (represented at different levels of abstraction) and the final implementation on a given platform. It also represents the information of the \textit{Context of use} using the \textit{Context Models}. To obtain this information, it uses a \textit{Monitor} (e.g., webcam, historic user data, light sensors).

    \item \textbf{Intelligent UI Adaptor}: is the component that detects changes in the context and makes the decision to adapt and perform the adaptation. 
    The \textit{Adaptation Manager} will perform the complete adaptation process from its initiation to its completion. It has its own UI (\textit{Adaptation Manager UI}) to enable the end-user to access and update the \textit{adaptation parameters}. It also has an \textit{Adaptation Engine}, which contains all the adaptation logic and an \textit{Adaptation Transitioner} which convey any transition from one UI to an adapted UI. 
    This component also has an \textit{Adaptation Explainer}, which justifies why is a specific adaptation being carried out. We expect that this component will enhance the users' trust with the framework.
    Finally, the \textit{Adaptation Agent}: it's using Reinforcement Learning algorithms (e.g., Q-Learning, SARSA, DQN), it will have to decide which of the multiple possible actions to adapt the UI (e.g. change to dark/light theme, change the layout of the UI, do nothing, among others) should be applied at any given time to improve the UX of the user. This agent is learning from user interaction since the process is iterative and evaluates the outcome of the actions that it has taken in the past. 

    \item \textbf{External Sources}: give more knowledge to the Intelligent UI adaptor to make more informed decisions. External sources include empirical data on user interaction as well as the real-time information gathered from user interaction.
    
    \item \textbf{Physiological Data}: it collects different aspects of the users' physiological and emotional state. It can be collected through various methods such as electroencephalography (EEG), eye tracking (ET), galvanic skin response (GSR), heart rate (HT), facial expression recognition, among others. Then, this data is processed to obtain different UX measures such as attractiveness, cognitive effort and user emotions (e.g., happy, sad, surprised).
\end{itemize}

In order to implement the intelligent UI adaptation framework, we first 
defined the UI adaptation problem as a stochastic sequential decision problem. We modeled it as a MDP, which allows us to represent sequential decisions in an uncertain environment. The MDP formulation consists of a set of \textit{states}, \textit{actions}, and a \textit{reward function} that guides the agent's learning process.

The \textit{state} space of our UI adaptation environment is defined by four subspaces: User Interface, Actor, Platform, and Environment. The User Interface subspace captures visual properties of the interface, including but not limited to: layout type (grid or list), theme (light or dark), font sizes and styling, widget types. This information can be obtained from the \textit{User Interface} component.
The \textit{Actor} subspace describes characteristics of the user, such as age, emotion, and previous experience. The \textit{Platform} subspace includes technical properties of the device (e.g., screen size, luminosity) and operating system used. Lastly, the \textit{Environment} subspace represents the physical location and ambient brightness. 
This multidimensional state space provides a comprehensive representation of the UI and its contextual factors. The \textit{Actor}, \textit{Platform} and \textit{Environment} information can be obtained by monitoring the \textit{Context of use} and its represented in the \textit{Contetxt Model}.

The agent (contained in the \textit{Intelligent UI adaptor}), that is being implemented using RL algorithms, interacts with the \textit{Software System} and performs adaptations by choosing \textit{actions} from a discrete action space. In our preliminary implementation, we have defined eight possible actions that allow the agent to modify the UI's layout (change from grid to list and vice-versa), color scheme (from dark to bright and vice-versa), and font size (change to small, default and big font sizes). Additionally, we have included a "not to adapt" action, enabling the agent to decide whether an adaptation is necessary in a given state. The agent's goal is to select actions that lead to UI adaptations maximizing the user's preferences and UX.

In RL-based UI adaptation, defining suitable rewards poses a challenge. However, we have developed a preliminary reward function that considers preference similarity and UX. Preference similarity measures the alignment between user preferences and adaptations, while UX evaluation incorporates usability (task completion time, success rate) and user emotions captured through a webcam. The reward function (Equation~\ref{eq:reward}) combines these criteria using weights ($w_i$) to guide the RL agent's decision-making process.
\begin{equation}
R = \sum_{i=1}^{4} w_i * c_i
\label{eq:reward}
\end{equation}

Nevertheless, we are exploring other approaches such as making the reward function part of the learning process. In complex environments, utilizing reward models, which are used to specify the users' goals and the system~\cite{Chen:2022}, can offer significant benefits. Human Feedback (HF) could be used as regards specifying the goal more intuitively and quickly when compared to manual objective hand-crafted methods~\cite{Armstrong:2021, openaiDeepmind:2017}.
Specifically, we are using predictive HCI, which are computational models that can explain how users interact with interfaces at the level of individual human cognition~\cite{Paton:2021}, to obtain the reward from each adaptation. It simulates consequences – benefits and costs – of possible adaptation sequences without actually executing them.

The proposed framework using OpenAI Gym as the RL agent training environment provides a flexible and extensible platform for experimenting with various RL algorithms, such as Q-Learning, SARSA, DQN, and A2C. Researchers and practitioners could use this environment to train and test different RL agents for UI adaptation in diverse platforms and usage environments.


\subsection{Evaluation Plan}
\label{sec:evaluationPlan}

To validate the effectiveness and impact of the proposed adaptive user interface framework, a comprehensive evaluation plan is devised. This plan incorporates both simulation-based experiments and user studies to gather empirical data and assess the performance of the framework in various contexts and scenarios.

In the simulation-based evaluation phase, simulations will be conducted using the reinforcement learning algorithm implemented within the AUI framework. These simulations will be designed to represent different user contexts and scenarios, allowing for the assessment of adaptation effectiveness, user engagement and user satisfaction. By creating a variety of simulated environments, the performance of the adaptive UI in terms of its ability to adapt to diverse user needs and preferences can be evaluated. Additionally, the results obtained from different RL algorithms will be compared to identify the most effective approach for UI adaptation.

User studies will be conducted to evaluate the user experience, by means of their user engagement, user satisfaction, and perceived usability of the adaptive UI. A diverse group of participants, representing different user profiles and preferences, will be recruited for the controlled studies. Controlled environments with specific tasks and scenarios will be created to assess the impact of the adaptive UI in real-world situations. User feedback will be collected to gain insights into the effectiveness and benefits of the proposed framework. We plan to use non-intrusive methods to obtain the different UX measurements, for example, by examining the user interaction patterns and using predictive HCI models~\cite{gaspar-figueiredo:2023dESEM}.

Another aspect of our ongoing research is exploring different ways of obtaining rewards to evaluate UI adaptations. As mentioned in the previous sections, the reward function plays an important role in reinforcement learning-based UI adaptation. We are investigating various reward models, including traditional objective hand-crafted methods, Human Feedback (HF)~\cite{gaspar-figueiredo:2023dESEM}, 
and potentially making the reward function itself a part of the learning process. Each approach has its advantages and limitations, and our goal is to identify the most effective and efficient way of evaluating adaptation quality for different contexts and user preferences.
%
%
Furthermore, we plan to do more experiments that include more human factors into the reward function such as the users' emotions. We plan to obtain the users' emotions through the analysis of the facial expressions using a webcam. We believe that this type of data can be used as non-intrusive feedback to give more insights to the RL agent.
%
%
%

This evaluation plan will provide valuable insights into the effectiveness, and impact of the proposed AUI framework. It will enable the identification of strengths, limitations, and areas of improvement, ensuring the continuous refinement and enhancement of the framework. 
The empirical evaluations with users will help validate the effectiveness in terms of the UX produced by adaptive UI in real-world scenarios.


\section{Methodology}
\label{sec:methodology}
In this section, we describe the process we are following to design and evaluate the framework for user interface adaptation based on the technology transfer model introduced by \cite{Gorschek:2006}. We used this model since our research method involved evaluations in both academia and industry with the aim of scaling the proposal up to practice, for which this model is recommended~\cite{wohlin:2012}.
The technology transfer model is a seven-step process for transferring research results from academia to industry. The steps include 
\textit{i) the problem formulation based on the the state-of-the-art} and \textit{ii) industry needs},
\textit{iii) formulation of a candidate solution}, \textit{iv), v), and vi) evolution and transfer preparation through validation in both academia and industry}, which include validation in academia, static and dynamic validation, and \textit{vii) releasing the solution}.

We first analyzed the state-of-the-art and industry needs.
We detected that there is a lack of approaches that adapt the whole user interface to improve the UX using RL. 
We then proposed a candidate solution (see Section~\ref{sec:propFramework}), which we are constantly evolving. This solution combines the use of RL and Physiological data to adapt the whole UI to improve the UX.
Afterwards we defined a study to validate the solution in the academia~\cite{gaspar-figueiredo:2023dESEM}. We want to validate as well the RL approach with different algorithms in a simulated environment.
Following the methodology, after the feedback obtained in the validation in academia, we will improve the solution and then we will validate it in the industry. Finally, after the validations, the solution will be released.

\section{Current Status}

This research started in October 2021. Since then we have already reviewed the state of the art, detected the problem, proposed a solution and started with experiments with human-participants to validate the process followed by the framework. By now, we are implementing and evaluating the solution to refine it until it's released. We plan to complete this research on September 2025.

As part of our research on external data sources, we conducted two experiments using EEG to capture users' physiological responses to different variations of graphical menus in the user interface. The experiments involved a diverse group of users, and EEG data was collected to measure attraction, cognitive load, engagement, and memorization levels~\cite{gaspar-figueiredo:2023aTosem, gaspar-figueiredo:2023bEASE}.
The analysis of the EEG data revealed distinct patterns and correlations related to the different graphical menu designs. Notably, there was a significant positive correlation between attraction measured using EEG and attraction measured using questionnaires, as well as between cognitive load measured using EEG and cognitive load measured using questionnaires. These findings suggests the potential of EEG data in providing valuable insights into users' attraction and cognitive load towards different interface variations.

These findings from the EEG experiments highlight the importance of incorporating physiological responses into the UI adaptation framework. The knowledge gained from these experiments serves as a foundation for making informed adaptations and enhancing the user experience in future iterations of the framework.

Moreover, we have made significant progress in developing the initial version of our framework~\cite{gaspar-figueiredo:2023c}, which is available on GitHub at \url{{https://github.com/ISSI-DSIC/UI-adaptation-RL-env}}. 

Additionally, we are expanding the state space representation to handle more complex UI designs, increasing the variety of adaptations in the action space, and refining the feedback mechanisms to provide accurate and non-intrusive evaluation of adaptation quality. 
We have defined a hand-crafted reward function and defined different predictive HCI models that can obtain the user engagement from the user interaction. 
This will enable the UI adaptor to make better, contextually appropriate adaptations, thereby improving the overall user experience.

\section{Future Work}
\label{sec:future}

One avenue for future work is the expansion and refinement of the proposed framework by expanding the state space representation, adding more actions to the action space and optimizing the reward mechanisms. This is not only an \textit{in progress} work but also a future work since we want the framework to continuously evolve. 

According to the Methodology proposed (Section \ref{sec:methodology}) after the \textit{Validation in Academia}, we will focus on the \textit{Validation in the industry}. By gathering feedback over extended periods and involving a broader user base, it will be possible to gain a comprehensive understanding of the impact of RL-based adaptive UIs on user experience.

Finally, we are examining the explainability aspect of the adaptation process, as it serves for building trust and acceptance among users. The \textit{Adaptation Explainer}, which is a component of our framework (Figure~\ref{fig:framework}), will be further developed and refined based on user feedback to strike the right balance between providing insights and not overwhelming users with constant justifications. Moreover, the \textit{Adaptation Transitioner} will be developed. Its purpose is to manage the transition from the original UI to its adapted version in a smooth and seamless manner. This transition plays an important role in ensuring a positive user experience by minimizing disruptions and maintaining continuity.


\begin{acks}

D. Gaspar-Figueiredo is recipient of a Predoctoral Research staff-training program (GVA ACIF/2021/172) funded by the GVA.
This work contributed to the project \textit{AKILA: User Interface Adaptation through User-Experience-based Reinforcement Learning}, funded by the GVA and ESF (CIAICO/2021/303).
\end{acks}

\bibliographystyle{ACM-Reference-Format}
\bibliography{main}


\end{document}